\documentclass[superscriptaddress,showpacs,aps,12pt,endfloats,byrevtex]{revtex4-1}
\usepackage[utf8]{inputenc}
\usepackage{amsmath,amssymb}
\usepackage{graphicx,textcomp}
\usepackage{color}
\usepackage{todonotes}
\usepackage{todonotes}
\usepackage[utf8]{inputenc}
\usepackage{braket}
\usepackage[normalem]{ulem}

\setcitestyle{numbers,square}

\definecolor{fcolor1}{RGB}{120,120,255}
\definecolor{fcolor2}{RGB}{0,0,255}

\definecolor{acolor1}{RGB}{255,120,120}
\definecolor{acolor2}{RGB}{255,0,0}

\newcommand{\comment}[1]{}

\begin{document}
\title{Elinvar effect in $\beta-$Ti simulated by on-the-fly trained moment tensor potential}

\author{Alexander V. Shapeev}
\affiliation{Skolkovo Institute of Science and Technology, Moscow, Russia}

\author{Evgeny V. Podryabinkin}
\affiliation{Skolkovo Institute of Science and Technology, Moscow, Russia}

\author{Konstantin Gubaev}
\affiliation{Skolkovo Institute of Science and Technology, Moscow, Russia}

\author{Ferenc Tasn\'adi}
\affiliation{Department of Physics, Chemistry, and Biology (IFM),
Link\"{o}ping University, SE-581 83, Link\"oping, Sweden.}

\author{Igor A. Abrikosov}
\affiliation{Department of Physics, Chemistry, and Biology (IFM),
Link\"{o}ping University, SE-581 83, Link\"oping, Sweden.}
\affiliation{Materials Modeling and Development Laboratory, NUST ``MISIS'', 119049 Moscow, Russia.}

\date{\today}

\begin{abstract}
A combination of quantum mechanics calculations with machine learning (ML) techniques can lead to
a paradigm shift in our ability to predict materials properties from first principles. Here we show
that on-the-fly training of an interatomic potential described through moment tensors provides the
same accuracy as state-of-the-art {\it ab inito} molecular dynamics in predicting high-temperature
elastic properties of materials with two orders of magnitude less computational effort. Using the
technique, we investigate high-temperature bcc phase of titanium and predict very weak, Elinvar,
temperature dependence of its elastic moduli, similar to the behavior of the so-called GUM Ti-based
alloys [T. Sato {\ it et al.}, Science {\bf 300}, 464 (2003)].
Given the fact that GUM alloys have
complex chemical compositions and operate at room temperature, Elinvar properties of elemental bcc-Ti
observed in the wide temperature interval 1100--1700 K is unique.
\end{abstract}
\maketitle

Theoretical predictions of materials properties have become increasingly more reliable with the
development of {\it ab inito} molecular dynamics (AIMD), which brings simulations conditions in accord
with the conditions at which materials exist in nature or operate in devices. Power of the approach has
been demonstrated in numerous applications, ranging from studies of materials at the Earth’s core
conditions \cite{vocadlo_possible_2003} to investigations of alloys for hard-coatings applications
\cite{shulumba_lattice_2016}.  Exploring finite-temperature effects via MD simulations allows one to
uncover exceptional effects at finite temperature, such as a collective superionic-like diffusion of
atoms in Fe \cite{belonoshko_stabilization_2017} and Ti \cite{sangiovanni_superioniclike_2019}.
However, AIMD calculations are too time-consuming, and this greatly limits their use in practice,
despite increasing demands in providing reliable theoretical data for the broad range of applications
in physics, chemistry, geosciences, materials science, and biology.  

In particular, first-principles simulations of elastic properties of materials at finite temperature
are becoming broadly relevant for fundamental research 
\cite{steinle-neumann_elasticity_2001} and knowledge-based materials design
\cite{de_jong_charting_2015,avery_predicting_2019}. However, their accurate predictions represent
an enormous computational task 
\cite{steneteg_temperature_2013,shulumba_temperature-dependent_2015}. One has to perform multiple
molecular dynamics calculations for relatively large, with hundreds of atoms, simulation cells applying
different distortions and explore the stress-strain relations as a function of temperature.
Moreover, going from a description of thermodynamic properties of materials at finite temperature to
simulations of their elastic constants the accuracy of calculations has to be improved by an order of
magnitude, from about 10 meV for potential energy differences to 1 meV. This puts tough requirements
on the technical parameters used in AIMD runs, as has been demonstrated, e.g. in studies of the
materials with strong elastic softening and low thermal expansion \cite{zhang_new_2015}. 

A traditional way of increasing efficiency of molecular dynamics simulations is to employ classical
interatomic potentials fitted to experiment and/or first-principles calculations. The drawback of
utilizing the (semi-)empirical interatomic potentials (or force-fields) with MD simulations instead
of performing AIMD is that off-the-shelf potentials typically do not have sufficient transferability,
and therefore their use for a descriptions of atomic configurations outside of the phase-space used
for the original fitting may be problematic. To overcome the issue, the so-called {\it learn-on-the-fly}
strategy has been suggested \cite{de_vita_novel_1997}. For example, augmentation of the classical
potentials via on-the-fly quantum mechanical calculations has been utilized to model brittle failure
in silicon \cite{csanyi_learn_2004}. However, the flexibility of such classical potentials is still
limited by their functional form.

Machine-learning (ML) of interatomic potentials (MLIP) have been put forward recently as an
alternative and highly promising approach what should combine high efficiency and sufficient accuracy
in solving relevant physical problems
 \cite{podryabinkin_active_2017,podryabinkin_accelerating_2019,chmiela_towards_2018,jinnouchi_--fly_2019,
PhysRevLett.122.225701}.
Unlike empirical potentials with a fixed functional form, their ML counterparts aim at approximating
an arbitrary interatomic interaction, e.g., by expanding it in a complete set of basis functions
whose arguments are local descriptors of atomic environments. In particular, moment tensors have
been introduced as descriptors \cite{shapeev_moment_2016}. The corresponding ML many-body interatomic
potentials have been called MTPs. They have been successfully applied in zero temperature
crystal structure prediction \cite{gubaev_accelerating_2019}, predicting thermal conductivity
of compounds \cite{korotaev_accessing_2019} or computing anharmonic free energy of refractory high
entropy alloys \cite{grabowski_ab_2019}. In the applications, the
MTPs are fitted on-the-fly to high-accuracy density functional theory (DFT) calculations and provide
{\it ab initio} level of accuracy in the description of thermodynamic properties of materials using
several orders of magnitude less computational resources than conventional AIMD simulations.
However, as has been specified above, calculations of elastic properties of materials typically
require order of magnitude more accurate calculations. To the best of our knowledge, applications
of MLIP have not been explored for such challenging tasks yet. 

In this study we employ MTPs in simulations of the elastic constants $C_{ij}$ of bcc-Ti
($\beta$-Ti) in the broad temperature interval, from 900 K to 1700 K. Ti is the base elements for
a broad family of technologically relevant alloys for airspace
\cite{peters_titanium_2003,pollock_alloy_2016},
nuclear energy \cite{zhang_long_2019}, hard-coatings \cite{bartosik_fracture_2017} and
biomedical \cite{geetha_ti_2009,chen_metallic_2015} applications.
From the fundamental point of view, bcc, or $\beta-$phase of Ti is highly challenging for
theoretical simulations because of its dynamical instability at zero temperature
\cite{hennig_classical_2008}. This means that bcc-Ti cannot exist at low temperature, but it
becomes stable not only dynamically, but also thermodynamically at temperatures above 1155 K (at ambient
pressure). Therefore, strictly speaking, any theoretical consideration of
bcc Ti and Ti-rich alloys requires finite-temperature simulations \cite{skripnyak_mixing_2020}. 

We demonstrate that our technique based on the use of MTPs predicts high-temperature elastic
properties of bcc-Ti with the same accuracy as state-of-the-art AIMD simulations, though with
two orders of magnitude less computational efforts. Importantly, in our simulations we
observe very negligible temperature dependence of elastic moduli of $\beta-$Ti in the wide
temperature interval 1100--1700 K. The Elinvar effect in bcc-Ti uncovered in our simulations
is similar to the elastic behavior of the so-called GUM alloy
\cite{saito_multifunctional_2003,wang_strain_2015,zhang_new_2015,talling_mechanism_2009}
with an important difference: while the latter have complex chemical compositions and operate
at room temperature, the former is a pure elemental metal showing remarkable properties at
very high temperatures.

In this work, $\beta$-Ti has been modeled with a supercell of 128 atoms. The exchange
correlation potential in all our VASP \cite{hafner_ab-initio_2008} calculations has been
approximated using PBE functional \cite{perdew_generalized_1996}. For AIMD simulations we
applied the energy cutoff 460 eV and a $\Gamma$-point centered (2$\times$2$\times$2)
k-point mesh. The elastic constants between 900 and 1700\,K have been calculated from the
stress-strain relations applying lattice distortions $\pm 1\%$ and $\pm 2\%$ and the NVT
corresponding stress-fluctuation formula. We performed 10\,000 AIMD steps with each distorted
supercells at each temperatures using the time step of 1 fs, see Ref.\cite{skripnyak_theoretical_2020}.
The averaging for MTP was done on 100 Metropolis-adjusted Langevin algorithm (MALA) \cite{roberts_optimal_1998} trajectories
with 10$^5$ steps for each distortions.

\begin{figure}[ht]
\includegraphics[width=8cm]{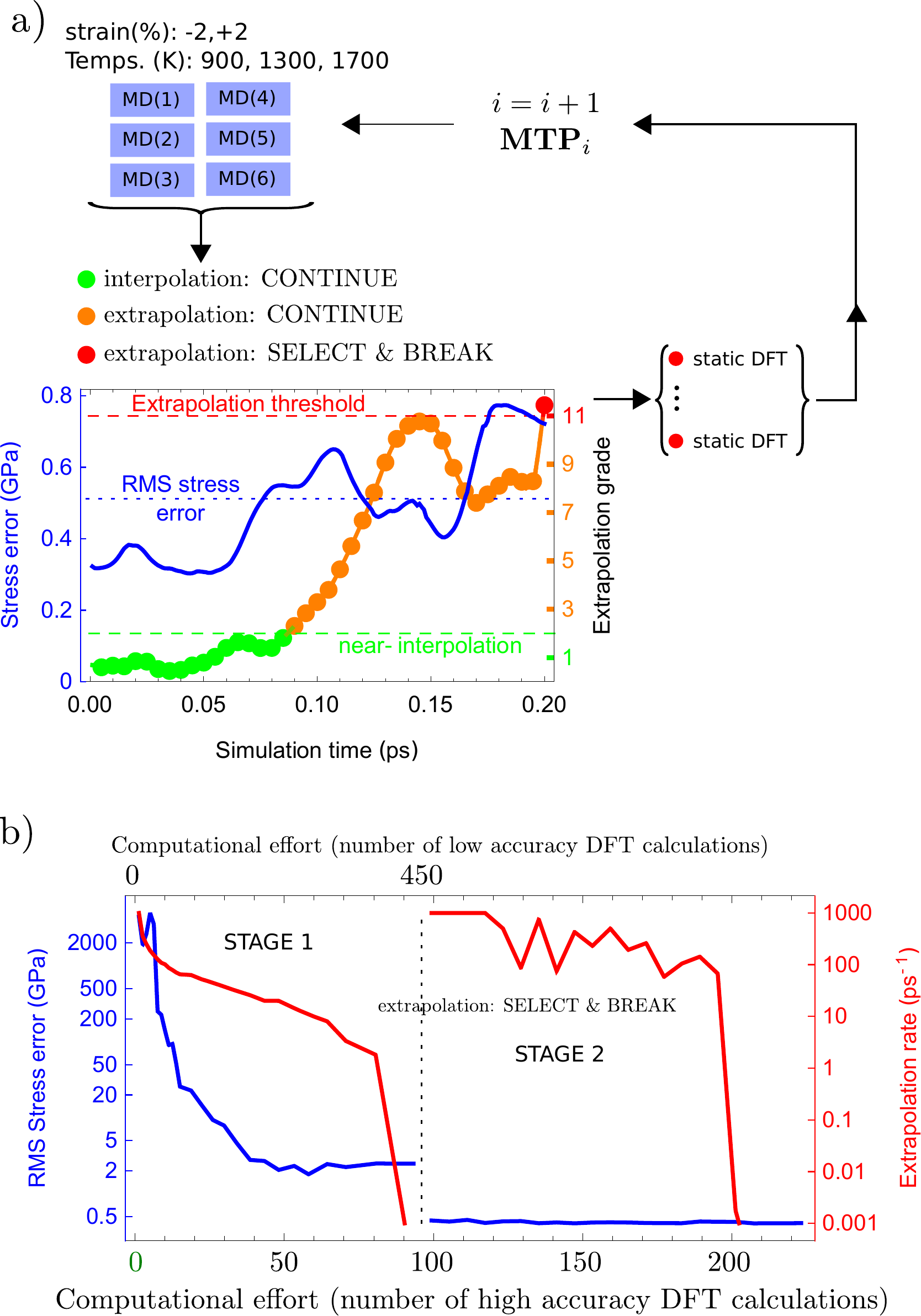}
	\caption{Our MLIP approach of training the MTP potential. a) displays the active learning process
	of building the training set (select, continue or break the MD simulations) and upgrading MTP.
	b) shows the convergence of both, the averaged root mean square (RMS) stress error and the averaged
	extrapolation grade wrt. the number of static DFT calculatons in our two-stage approach.}
\label{fig_01}
\end{figure}

In our MLIP approach we have used MTP built from polynomial-like
basis functions of degree 16 or less, see Ref.\cite{gubaev_accelerating_2019}, which resulted
in about 120 parameters to fit. Our two-stage active-learning process is illustrated in
Fig.\,\ref{fig_01}. The initial training, see Fig.\,\ref{fig_01} a), set contained six (three temperatures and two 
different strains) ideal atomic configurations with the experimental bcc lattice parameters
\cite{thurnay_thermal_1998,senkov_effect_2001,spreadborough_measurement_1959}
at the corresponding temperature. For the training set we sampled configurations at
900, 1300 and 1700 K and applied $\pm\,2\%$
strains on the simulation box. We performed the corresponding six MD simulations in parallel
using LAMMPS \cite{plimpton_fast_1995}. The extrapolation grade of the MTP in each MD at
each time step was calculated by the MaxVol algorithm \cite{olshevsky_matrix_2010}.
The arithmetic average of the (six) extrapolation rates is shown in Fig.\,\ref{fig_01}

An MD was stopped at a configuration whose the grade exceeded the extrapolation threshold,
Fig.\,\ref{fig_01} a), 
and static DFT energy calculation was performed on that extrapolative configuration.
Then the training set was expanded the extrapolative configurations (we used at most six configurations)
and the MTP was retrained. This process was repeated until the extrapolation rate
(inverse of the reliable simulation time) reached zero for each MD. During stage 1 we
used VASP and Projector Augmented Wave (PAW) potentials \cite{blochl_projector_1994} with only four
valence electrons. This stage resulted in 450 configurations and 3 GPa RMS stress error of
the MTP on the training set.

At the second stage the goal was to fit a potential to the high-accuracy DFT calculations
with 12 valence electrons PAW potential of Ti. In the beginning the converged
first stage MTP was utilized to sample 100 configurations with Boltzmann statistics and
the total energies and forces for each of the configurations were subsequently computed
with DFT. At this point all the low-accuracy DFT calculations were discarded and the MTP
was fitted to the 100 accurate DFT calculations. In the continuation, similar to our approach
at the first stage, parallel MD simulations were repeated until achieving a vanishing
extrapolation rate. About 20 accurate DFT calculations were done at the second stage,
in addition to the initial 100 configurations, to obtain the final MTP potential. The converged
MTP resulted in fitting error of 2.2 meV/atom
for energy differences, 97 meV/\AA\ absolute error in forces, and 0.3 GPa absolute
error in stresses. Based on the average computational core-hours of the high-accuracy DFT
calculations the first stage 450 low-accuracy DFT calculations are estimated to correspond
to about 100 high-accuracy DFT calculations, as it is plotted in Fig.\,\ref{fig_01} b).
Thus, we estimate the computational efforts required for the full determination of our MTP potential
to correspond to $\sim$200  accurate DFT calculations. 

\begin{figure}[ht]
\includegraphics[width=8cm]{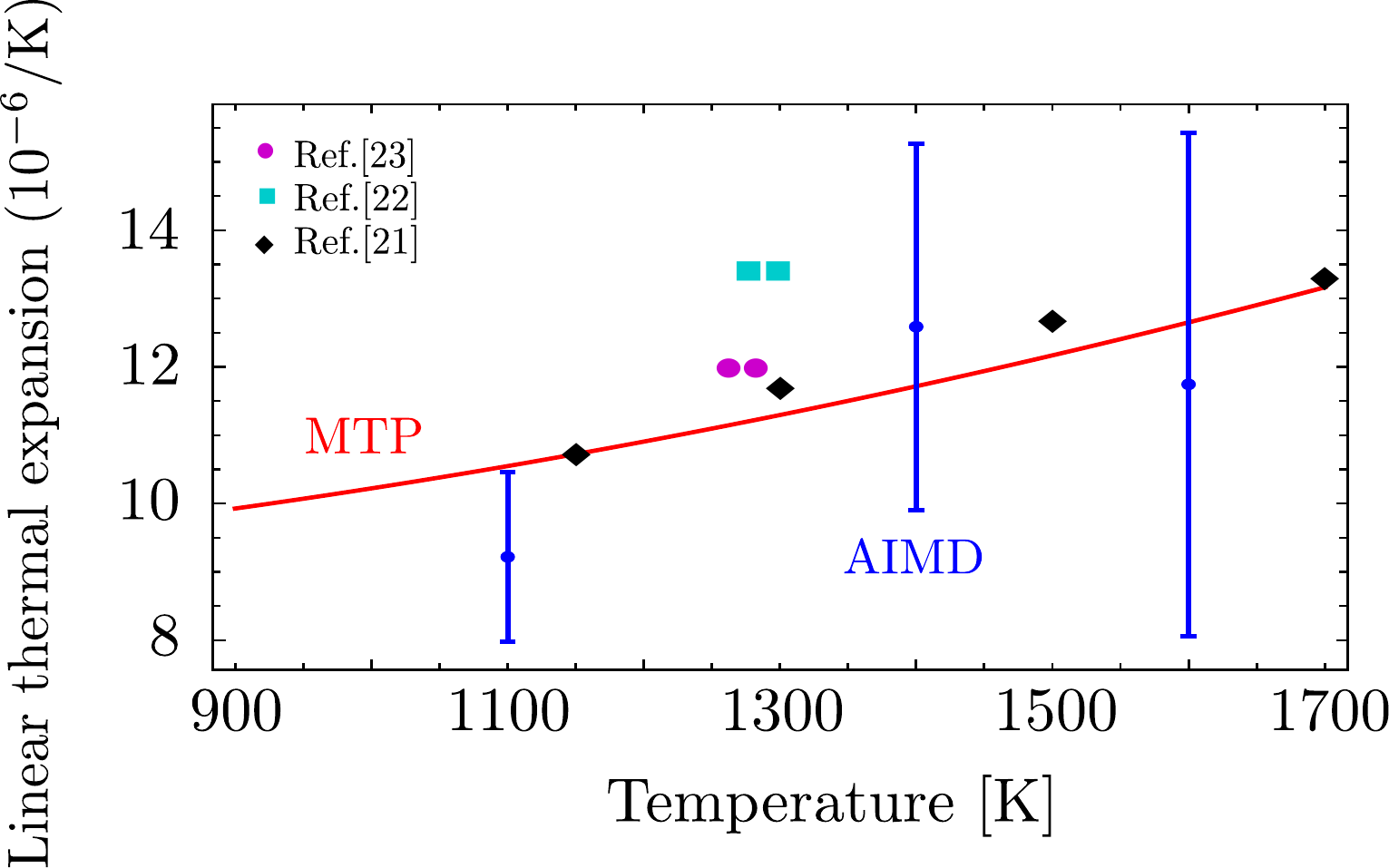}
\caption{Thermal expansion of $\beta$-Ti obtained by MTP in comparison with results derived by
	AIMD simulations and experimental data.}
\label{fig_02}
\end{figure}

Let us next verify the accuracy of our potentials in a comparison with  available experiemtal information. 
Fig.\,\ref{fig_02} shows our derived thermal expansion coefficients of $\beta$-Ti in comparison with
experimental values from Refs.
\cite{thurnay_thermal_1998,senkov_effect_2001,spreadborough_measurement_1959}.
The results obtained 
with MTP are given as a continuous line. In addition, we show results of direct AIMD calculations at
1100, 1400 and 1600\,K. Finite difference approach was applied for the estimation of the
temperature derivatives. Both, direct AIMD calculations of the thermal expansion
and simulations with MTP result in good agreement with experiment. The estimated linear increase of
the lattice parameter between 900 and 1700 K is 0.8\%. Though the increase is relatively small,
it is in the range what is expected for metals. The large error bars of AIMD calculations relative
to the actual thermal expansion illustrate an advantage that machine-learning potentials
bring: with just a few hundred static DFT calculations they offer excellent accuracy despite the
small expansion coefficient and dynamical instability of bcc-Ti at zero temperature, both of
which have adverse effects on the statistical error of the AIMD method.

\begin{figure}[ht]
\includegraphics[width=8cm]{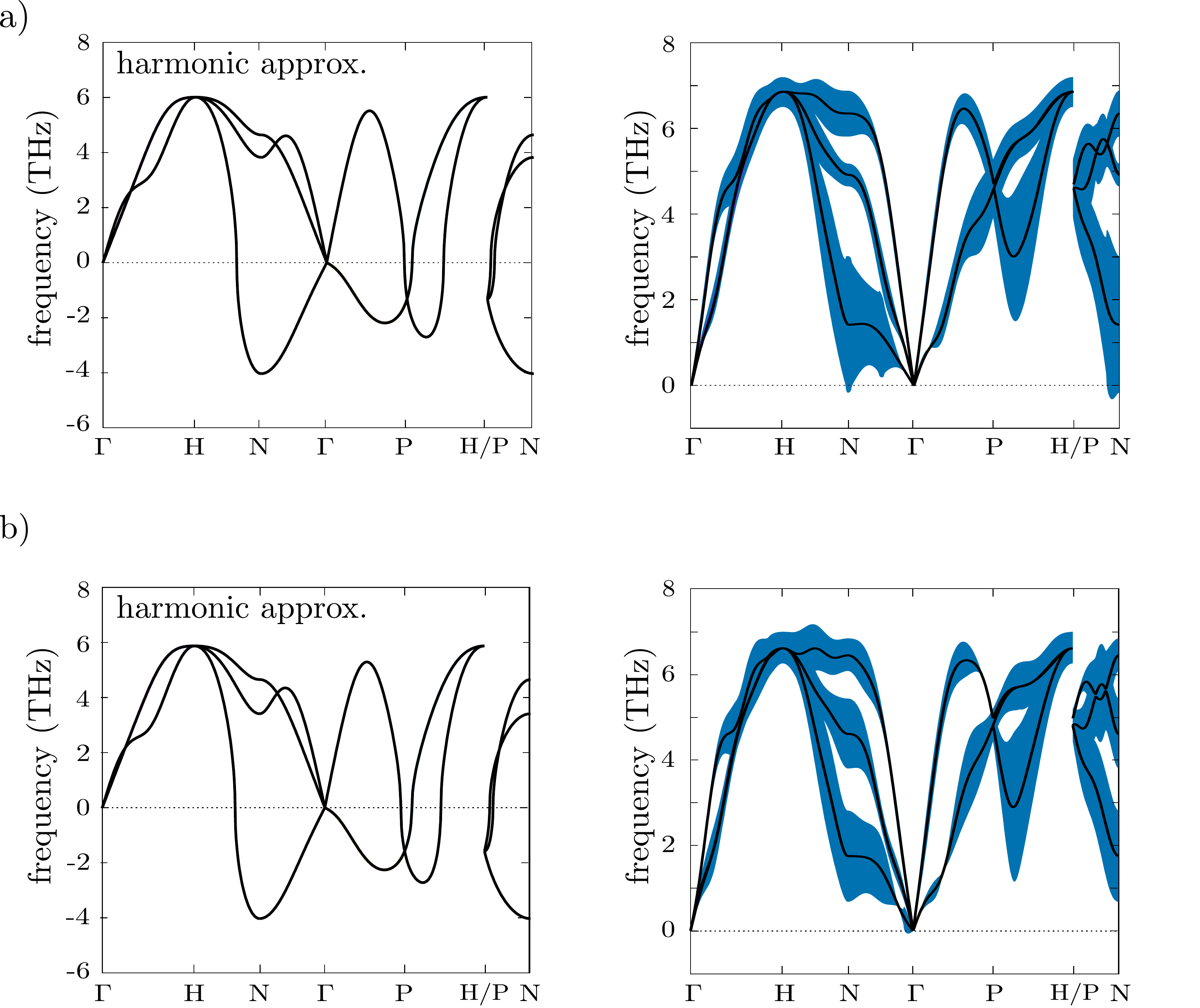}
	\caption{Phonon dispersion relations in $\beta$-Ti at a) 900\,K and
	b) 1500\,K using the derived MTP. The left panels show phonon dispersions within the quasiharmonic
	approximation using finite displacements with MTP.}
\label{fig_03}
\end{figure}

Next, we utilized our MTP to obtain phonon dispersion relations in $\beta$-Ti calculating velocity
autocorrelation function and using normal-mode-decomposition implemented into
DynaPhoPy \cite{carreras_dynaphopy:_2017}. Figure \ref{fig_03} shows the derived phonon dispersions
and linewidths extracted from the force constants of a $(6 \times 6\times 6)$ supercell and velocity
autocorrelation functions calculated with MD simulations of a a $(12 \times 12\times 12)$ supercell
at 900 and 1500\,K. 

Quasiharmonic phonons are calculated using the finite displacement approach.
Assumption of a harmonic Hamiltonian
with force constants derived from the obtained MTP 
combined with the thermal expansion of the lattice shown in Fig.\,\ref{fig_02}
results in the well-known strong dynamical
instability of bcc Ti \cite{ko_development_2015} around N-point and between P an H points.
The corresponding large phonon linewidths (or short lifetimes) are shown with the shaded areas
in Fig.\,\ref{fig_03}. It is known that the N-point instability promotes the bcc-hcp transition
\cite{liu_bcc--hcp_2009} while the other one is connected to the formation of the $\omega$ phase. 
On the contrary, the phonon dispersions relations calculated from full MD with MTP potentials show
the dynamical stability of bcc Ti at high temperature, and  are in agreement with experiment and with
calculated data from literature \cite{sangiovanni_superioniclike_2019,ko_development_2015}.
Comparison of the results calculated at 900 and 1500\,K points
to an interesting effect: the broadening decreases with temperature, which is quite unusual. The effect
might be related to the fact that the bcc phase of Ti has barely became stable dynamically (and it is
still unstable thermodynamically) at 900, while it is becoming increasingly stable with temperature increasing
to 1500\,K. The effect can also be seen as a "precursor" showing that the temperature dependence
of physical properties in bcc Ti might be anomalous. Let us demonstrate that this is indeed the case
considering temperature dependence of elastic properties of $\beta$-Ti.

\begin{figure}[ht]
\includegraphics[width=8cm]{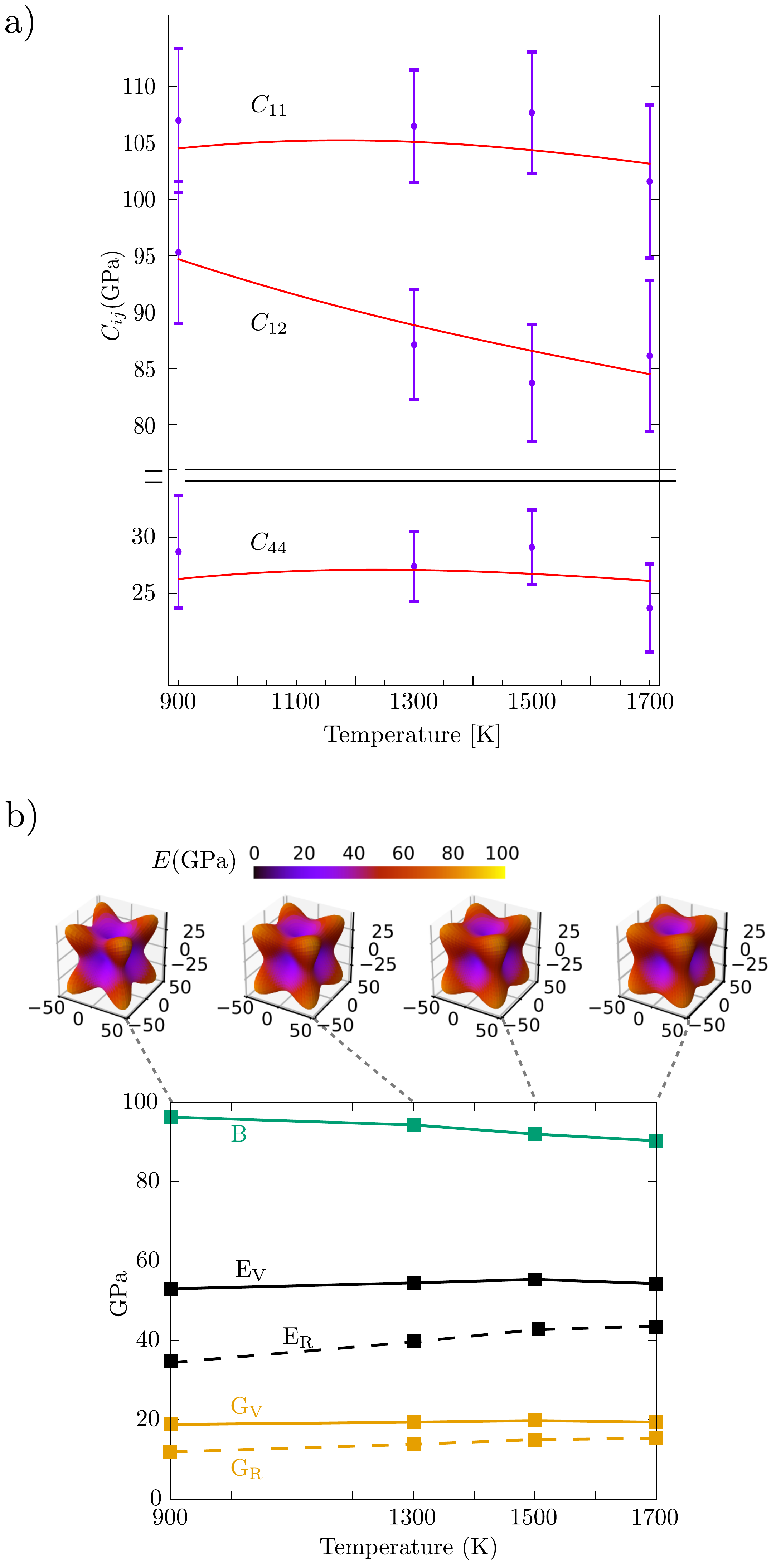}
	\caption{a) Temperature variation of elastic stiffness constants $C_{ij}$ of $\beta$-Ti.
	b) The directional variation of Young's modulus and polycrystalline bulk (B), shear
	(G) and Young's (E) moduli of $\beta$-Ti using Voigt (V) and Reuss (R) averages
	at temperatures 900, 1300, 1500 and 1700\,K.}
\label{fig_04}
\end{figure}

The calculated elastic constants and polycrystalline moduli are presented in Fig.\,\ref{fig_04}.
MTP predicts $C_{ij}$s with negligible statistical error and in excellent agreement with results of AIMD
simulations \cite{skripnyak_theoretical_2020}. Note that because of the harder convergence criteria, the
220 static DFT calculations used for training MTP in terms of the computational time corresponds to about
1000 AIMD time steps. Thus, we estimate that our scheme allows for over two orders of magnitude speedup
compared to calculating elastic constants from AIMD with the same accuracy and reliability of the
calculated results. 
Analyzing the calculated elastic moduli, $C_{12}$ decreases by just $\approx$ 7 GPa in the 
temperature range of the experimental thermodynamic stability of the bcc phase, between 1100 and 1700 K,
while $C_{11}$ and $C_{44}$ are almost independent on the temperature. Importantly, the calculated
polycrystalline elastic moduli, the bulk (B), Voigt (V) and Reuss (R) averaged shear (G) and Young's (E)
moduli, that are more relevant for the materials performance in realistic applications in comparison
to the single-crystal elastic moduli, all show negligible variation between 1100 and 1700\,K.
The observed behavior of the temperature dependence of the elastic moduli allows
us to argue that $\beta$-Ti manifests Elinvar effect.

In experimentally synthesized GUM alloys the Elinvar behavior is explained through a strain glass state
of the material with high defect concentration \cite{wang_strain_2015}. Random nano-domains of parent
and martensite phases contribute opposite to elasticity, which results in that matreial's elastic
moduli do not decrease with
temperature. Interestingly, Elinvar effect in $\beta$-Ti predicted in our simulations
is an intrinsic property of the material. Giving the fact that GUM alloys have complex chemical
compositions the Elinvar property of elemental bcc-Ti observed in the wide temperature interval
1100--1700 K is unique.

In summary, we have shown that an efficient on-the-fly machine
learning approach can be utilized to predict finite-temperature elastic 
behavior of materials with around two orders of magnitude less 
computational efforts than by conventional AIMD
simulations. The obtained results, such as thermal expansion and phonon 
dispersion of the high temperature bcc phase of titanium are in
agreement with available experiments. Furthermore, we have observed negligible 
temperature variation of the elastic moduli in $\beta$-Ti. This prediction open
up an opportunity for a
realization of the Elinvar effect observed in GUM alloys in an elemental metal
at high temperatures. Our results 
demonstrate a genuine and accurate approach to explore materials with useful
high temperature engineering properties.


Support provided by the Knut and Alice Wallenberg Foundation
(Wallenberg Scholar Grant No. KAW-2018.0194), the VINN Excellence Center Functional Nanoscale
Materials (FunMat-2) Grant 2016-05156 and the Swedish Government
Strategic Research Areas in Materials Science on Functional Materials at Linköping University
(Faculty Grant SFO-Mat-LiU No. 2009-00971) is gratefully acknowledged. Theoretical analysis of the
elastic properties was supported by the Ministry of Science and High Education of the Russian
Federation in the framework of the Increase Competitiveness Program of NUST “MISIS”
(Grant No. K2-2019-001) implemented by a governmental decree dated 16 March 2013, No. 211.
Development of the MLIP potentials was supported by RFBR grant number 20-53-12012.
The simulations were performed on
resources provided by the Swedish National Infrastructure for
Computing (SNIC) at the PDC Center for High Performance Computing at the KTH Royal Institute of
Technology and at the National Supercomputer Centre at Link\"oping University.
%
\bibliography{bcc-Ti}
\end{document}